\def\apj{{\it ApJ}}
\def\jcap{{\it JCAP}}

\def\aj{{\it Astronomical. J}}
\def\apjs{{\it ApJ. S}}
\def\nat{{\it Nature}}

\def\apjl{ApJL}
\def\mnras{MNRAS}
\def\aap{A\&A}

\def\ltsim{\lower.5ex\hbox{$\; \buildrel < \over \sim \;$}}
\def\gtsim{\lower.5ex\hbox{$\; \buildrel > \over \sim \;$}}
\def\ltsim{\lower.5ex\hbox{$\; \buildrel < \over \sim \;$}}
\def\gtsim{\lower.5ex\hbox{$\; \buildrel > \over \sim \;$}}

\def\vv{{\bf v}}

\def\vq{{\bf q}}
\def\kms{\mbox{km\,s$^{-1}$}}

\def\vB{{\bf B}}
\def\vx{{\bf x}}

\def\msun{{M_\odot}}
\def\vpsi{\bm{\Psi}}
\def\vnabla{\bm{\nabla}}
\def\hmpc{\ {\rm h^{-1}Mpc}}
\documentclass{iau}
\usepackage{amsmath}
\usepackage{graphicx}

\usepackage{natbib}

\usepackage{bm}

\title[Dynamics of the Large scale structure ] 
{Dynamics of The Tranquil Cosmic Web}

\author[A. Nusser]   
{Adi Nusser
}

\affiliation{Technion- Israel Institute of Technology \\ 32000
Haifa, Israel\\ email: {\tt adi@physics.technion.ac.il}} 

\pubyear{2014}
\volume{308}  
\pagerange{119--126}
\jname{The Zeldovich Universe: Genesis and Growth of the Cosmic Web}
\editors{R. van de Weygaert, S. Shandarin \& J. Einasto, eds.}
\begin{document}

\maketitle

\begin{abstract}
The phase space distribution  of matter  out to  $\sim 100 \rm Mpc$  is probed by two types of observational data: galaxy redshift surveys  and peculiar motions of galaxies.
Important information on  the process of  structure formation and deviations from standard gravity  have been  extracted from  the accumulating   data.
 The remarkably simple Zel'dovich approximation is the basis for much of our  insight into the dynamics of structure formation and the development of data analyses methods.
 Progress in the methodology and some recent results is reviewed.  

\end{abstract}

\firstsection 
\section{Introduction}

\begin{figure}[b]
 \vspace*{0cm}
\begin{center}
 \includegraphics[width=5.3in]{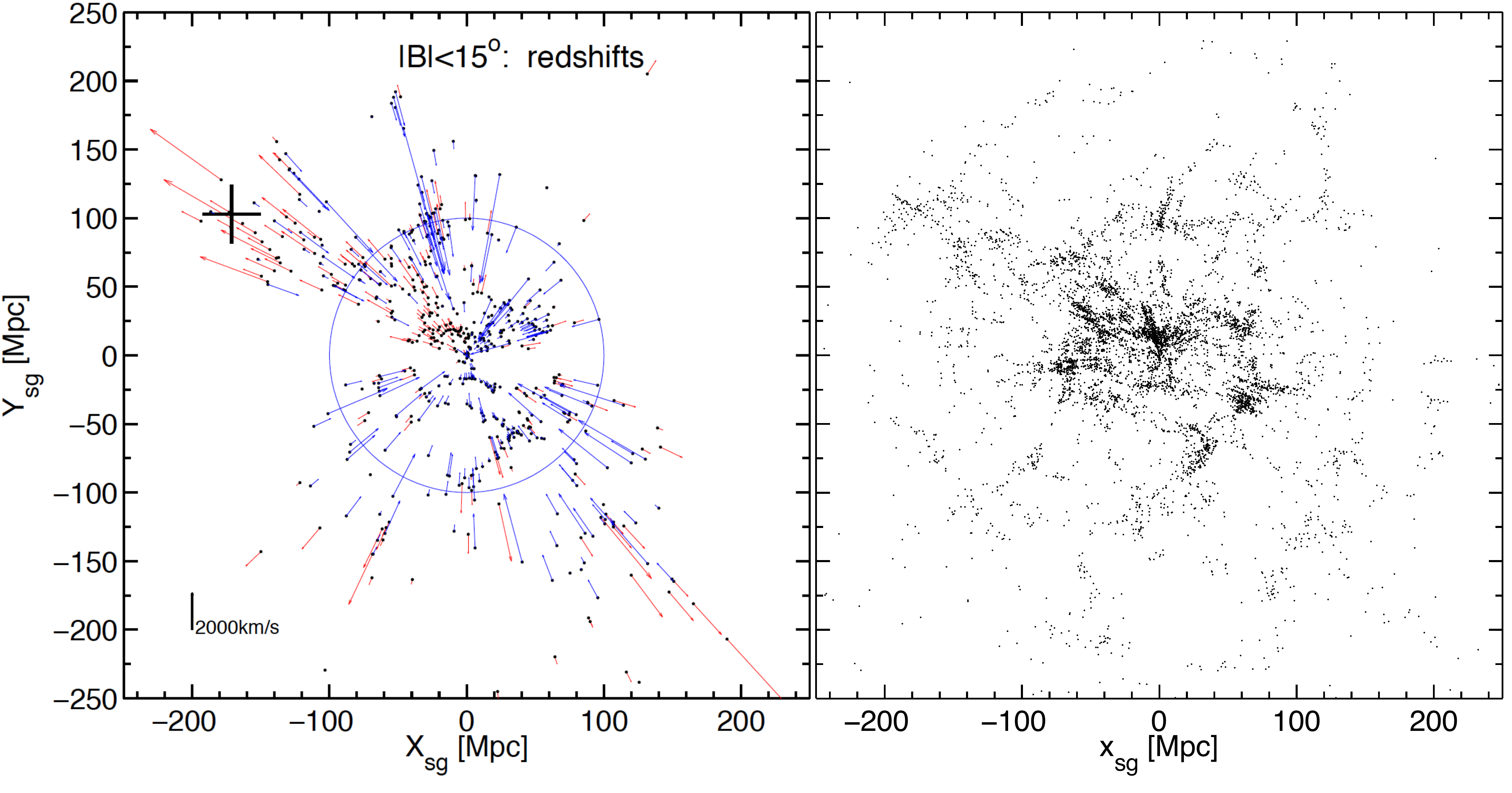} 
 \vspace*{-.5 cm}
 \caption{{\it Left: }The measured $v_p$ (in the CMB frame) of   galaxies within $15^\circ$
 of the SG plane in  CF2. Black dots indicate the observed redshifts in the CMB frame. Red and blue arrows correspond to 
 peculiar motions pointing away and towards the observer, respectively. 
The large cross plus sign indicates the location of the Shapley supercluster.
   {\it Right:} The distribution of galaxies within 20Mpc of the SG plane.  }
   \label{fig1}
\end{center}
\end{figure}

Merging and star formation activities in the galaxy population have  calmed gown by the current epoch ($z=0$). This lead to the establishment 
of  $a)$ 
a tight relation between 
the distributions of galaxies and the underlying mass  of the dark matter, and $b) $  relations 
between galaxy intrinsic properties, allowing for measurements of distance. 
Therefore, the $z\sim 0$ Large Scale Structure  is an excellent laboratory for 
probing cosmological models. 
Two complementary observational sets are our main window to the phase space distribution  of matter. The first, surveys of galaxy redshifts, $cz$,  
and apparent magnitudes, $m$. The second,  distance measurements $d_e$,  and hence peculiar motions
$v_p$, of galaxies obtained via intrinsic relations such as the Tully-Fisher (TF).   Distance measurements are  more difficult to obtain
than just $cz$ and $m$ and hence the number of galaxies with observed peculiar motions is significantly smaller 
than in redshift surveys. 
An example of the first set is Two Micron All Sky Redshift Survey  (2MRS) \citep{huchra12}, of about 45000 galaxies with a mean redshift 
of $\sim 8000\rm km s^{-1}$ and the deeper  SDSS containing about half a million galaxies  but with partial  sky coverage.
The second type of data include  the SFI++ catalog \citep{mas06} of TF measurements of $\sim 4000$ galaxies, and the Cosmic Flows 2 (CF2)\citep{CFTWO} catalog of $\sim 8000$.  
Fig. \ref{fig1} is a visual representation of the  data in the Super-galactic (SG) plane. Note the patchiness and sparseness of galaxies in the CF2 catalogue (left) compared to the 2MRS (right). 

\begin{figure}[b]
 \vspace*{0cm}
\begin{center}
 \includegraphics[width=5.in]{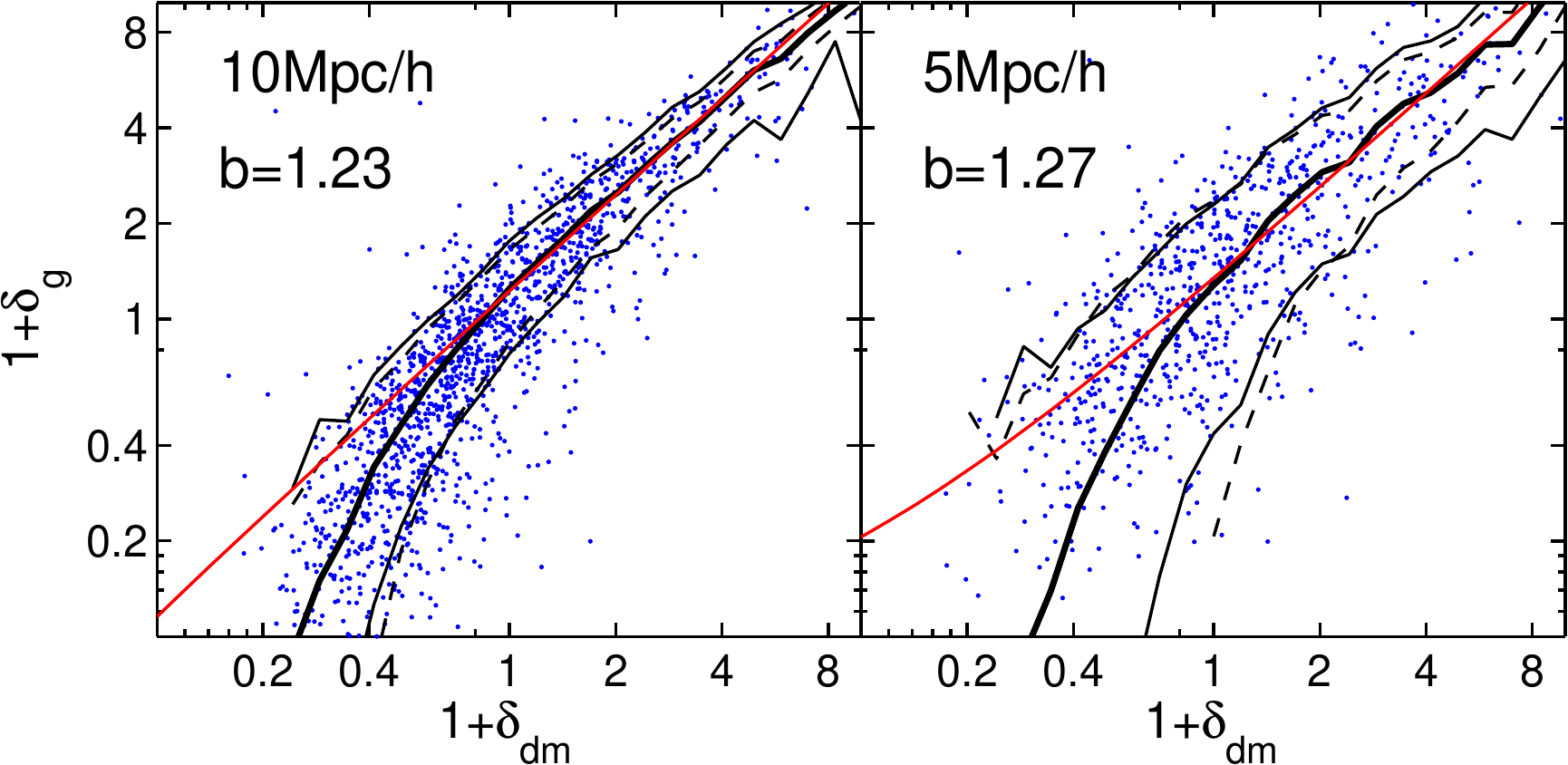} 
 \vspace*{0 cm}
 \caption{A scatter plot (logarithmic scale) of the galaxy  versus the dark matter over-densities
in  2MRS mock galaxy catalogs \cite{delucia}. The left and right panels correspond to  densities in cubic cells 
of $10\hmpc$ and $5\hmpc$ on the side, respectively.
 The thick solid curve in each panel is the mean of $1+\delta_{\rm g}$ 
at a given $1+\delta_{\rm dm}$. The two thin solid curves are  $\pm 1\sigma $ scatter computed from points above and 
below the mean.  Dashed curves are the expected $\pm 1\sigma$ Poisson (shot-noise) scatter.
The nearly straight red lines show $\delta_{\rm g}=b\delta_{\rm dm}+const$, where $b$ (indicated in the figure) are determined using linear regression  from 
 points in the range $-0.5<\delta_{\rm dm}<4$.  
     }
   \label{fig2}
\end{center}
\end{figure}

While the galaxy distribution is a biased tracer of the underlying mass density field of the dominant dark matter, the equivalence principle implies that galaxies are
comoving with the dark matter on large scales away from non-gravitational forces. But  
the peculiar velocity field (as a function of the measured distance, $d_e$) derived from the noisy data suffers from inhomogeneous  Malmquist biases \citep{lyn88}, resulting from the systematic difference between 
$d_e$ and the mean of true distances of galaxies with the same   $d_e$. 
It is very difficult to correct for this bias because of its dependence on  the unknown 
 distribution of galaxies in true distance space. In contrast,  
galaxy biasing  is likely to be well approximated by a simple  linear relation $\delta_{gal}=b \delta_{dm}$ between the 
galaxy and dark matter density fluctuations, as   seen in Fig. \ref{fig2}.

In the standard paradigm, the observed structure has grown by Gravitational Instability (GI) from tiny 
initial fluctuations. 
 Neglecting gas related effects, 
 the equations of motion (EoM) of the perturbations are the usual  Euler, Poisson and continuity  equations  in an expanding background.
%
Supplemented with initial conditions appropriate for cosmological perturbations, the full solution to the EoM
 is possible only via numerical  simulations which have achieved 
an impressive dynamical range from  small galaxies to a significant fraction of the Hubble volume. 
Nonetheless, approximate solutions have been and will remain the basis for  
observational analyses methods and 
a physical understanding the numerical   results.
The simplest approximate solution is provided by  linear theory  which yields
$\delta(\vx,t)=\delta_0(\vx) D+(t)+\delta_-(\vx)D^-(t) $ where $D^-$ describes a decaying mode and 
the growing mode obeys $\ddot D +2H\dot D-3\Omega H^2 D/2=0$. Linear theory also yields 
\begin{equation}
\label{eq:lin}
\delta=-\frac{1}{H f}\vnabla \cdot \vv\; ,
\end{equation} 
where $f=d\ln D/d\ln a \approx \Omega^\gamma$ is the growth rate and $\vv(\vx) $ is the 3D peculiar velocity field. The index $\gamma $ depends on the cosmology (e.g. through the dark energy model) and the underlying theory for gravity. Accurate determination of $\gamma$ is one of the goals of  future large surveys of galaxies. This $\delta-\vv$  has been used extensively for the prediction of velocity fields associated with the distribution of 
galaxies in a given redshift survey. It is the basis for modeling redshift distortions of 
correlation functions from redshift surveys on large scales. 

Within GI the two independent data sets  can be analyzed in several ways:
$a) $  Correlation functions (and power spectra) have been estimated from the distribution of galaxies in redshift surveys. 
These correlations can be compared with predictions of cosmological models. Further, $cz=Hd +v_p$ implies that 
correlations in redshift space indirectly probe $v_p$ through the fingers of god effect on small scales and the enhancement 
of clustering on large scales \citep{DP83,k88}.
$b)$ Correlation analysis of the observed peculiar velocity field have also been done. However, this analysis is intrinsically  plagued with 
 inhomogeneous Malmquist biases. Quantitative conclusions from this type of analysis should always be examined 
critically. 
$c)$ Comparison  of low order moments of the peculiar velocity field, e.g. the { bulk flow}, with predictions of cosmological models. 
$d)$ Testing  GI by assessing the alignment  of the gravitational force field (or the peculiar velocity) derived from redshift survey with the observed ${v_p}$ of galaxies in the peculiar velocity catalogs. 
This comparison is particularly important  since it minimizes cosmic variance in the estimation of the cosmological parameters.

\section{The Zel'dovich approximation} 

Full analytic solutions to the EoM are available only for initial conditions with a high degree of 
 symmetry, e.g. self-similar collapse/expansion with planar, cylindrical or
 spherical symmetry.
 \cite{zeld70}  proposed a remarkably simple   approximation for  the evolution of generic cosmological 
 perturbations in the quasi-linear regime (laminar flow).  The Zel'dovich approximation (ZA) states that the  current position $\vx(t)$ and the initial Lagrangian coordinate, $\vq$, of a particle are related by
 \begin{equation}
 \label{eq:zeld}
 \vx=\vq+D(t){\boldsymbol \psi}(\vq)\; .
 \end{equation}
  In  the  paper, Zel'dovich considered only baryons and   argued that the natural perturbation scale is  the Silk damping mass scale, $M_{\rm S}\approx 10^{12}M_\odot$. 
  Further, the  probability distribution of  the eigenvalues of $\partial_i \psi_j$, revealed a preference for planar-like 
  perturbations. Hence,  ZA was the basis for the top-down pancake paradigm for 
  structure formation.
 The approximation (\ref{eq:zeld}) is an exact solution to the full EoM for planar  perturbations 
 up to the onset of shell crossing (in collision-less fluids). The proof appears in 
 \cite{1983reas.book.....Z},  but not in the 1970 paper. 
 Although not highly accurate the ZA has given us fantastic physical insight into the working of nonlinear dynamics, 
 e.g. the growth of angular momentum of galaxies \citep{DOROSH70,White84}, nonlinear density power spectrum with and without redshift distortions  \citep{1995MNRAS.273..475S,FN96,TH96,2014MNRAS.439.3630W}  and the probability 
 distribution of the density field \cite{kofman}.

 \subsection{Extension: Lagrangian Perturbation Theory}
Here, the displacement is expanded in a Taylor series in an appropriate  parameter \citep{1993MNRAS.264..375B} which can be taken as the linear growth factor, $D$. 
Therefore, 
   $\vx=\vq+\sum_s D^s \vpsi^{(s)}(\vq)$, where 
   the ZA term ($s=1$) is entirely fixed by the initial conditions, while the EoM  dictate  all $s>1$ terms
via a  recurrence relation involving lower order terms only \citep{2014JFM...749..404Z}.
 Unfortunately these recurrence relations become messy for $s>2$. One of the reasons for that is 
 the  emergence of non-vanishing  Lagrangian vorticity  $\vnabla_q\times  \vpsi^{(s)}\ne 0$ for $s>2$ \footnote{A vanishing Eulerian vorticity as a function of 
 time 
 (for an initial irrotational flow) is protected by Kelvin's circulation theorem until the onset of orbit-mixing. In other words, Eulerian vorticity remains zero for any order in perturbation theory.}.
 Second order Lagrangian perturbation (2LPT)  ($s=2$),  
gives the density as 
 \begin{equation}
\frac{1}{\rho_{_{\rm 2LPT}}}
= 1+D  \vnabla \cdot \vpsi^{(1)} +{\frac{4}{7}}D^2(\mu_1 \mu_2+\mu_1 \mu_3+\mu_2\mu_3)\; .
 \end{equation}
 where $\mu_i$ are the eigenvalues of $\partial_i \psi^{(1)}_j$.
The continuity equation yields the widely known  expression for the density in the ZA, 
\begin{equation}
\frac{1}{\rho_{_{\rm zel}}}= 1+D  \vnabla \cdot \vpsi^{(1)}  +D^2(\mu_1 \mu_2+\mu_1 \mu_3+\mu_2\mu_3)+D^3 \mu_1\mu_2\mu_3\; .
 \end{equation}
 The ZA and 2LPT share the algebraic form of  the second order term ($D^2$),  but with  different coefficients. Thus,  ZA is inaccurate even to second order. However, simulations show that 
The ZA provides a  better match to the density derived from the velocity in 
 \citep{Gramann} in high density regions,  although 2LPT is better for low densities.
 
One of the most important common application of these approximation is the generation of particle displacements and velocities to be used as initial conditions for N-body simulations. The ZA and 2LPT are traditionally used but the latter yields is more accurate for this purpose.

\section{Peeble's action method} 
We are given the positions  of mass tracers (galaxies) today. What  are the tracers' paths  
from the nearly homogeneous early Universe to the current configuration? This is {\it a boundary value problem} where  a solution to the equations of 
motion is sought for {\it boundary conditions}  (BC) at two different times\footnote{It is closely tied to transport problems where  displacements from a clumpy into a uniform distribution are  sought by minimizing a cost function \citep{f02}. One still needs a dynamical prescription, e.g. ZA,  to get the orbits from total displacements. }.
Its solution allows a reconstruction of the  velocity field associate with the observed distribution of tracers and also 
the initial density field which lead to this distribution through gravitational interactions. 

ZA and 2LPT can be employed to derive approximate solutions to the orbits. However, the most general (and elegant) method to do that  has been designed by  \cite{peeb89} based on the least action principle.  
Orbits, $x(t)$,  obeying the EoM also render  the action stationary  with respect to variations, $\delta x(t)$, satisfying 
the BC
 $p\delta x=0$ at the limiting times $t_1$ and $t_2>t_1$, where $p$ is the momentum. 
In the cosmological problem,  $\delta x(t_2)=0$ is naturally  imposed. The initial positions are unknown, but Peebles noted that for the growing mode of cosmological perturbations the momentum vanishes as
$t_1\rightarrow 0$. Hence,   the solutions to the cosmological boundary value problem can be obtained by minimizing the action with respect to 
trial orbits constructed to satisfy  $a)$ known current 
positions and $b)$  $ p\rightarrow 0 $ near the Big Bang. For sufficiently general trial functions the orbits should be a solution to the full equations satisfying the BC, should such a solutions exist.
In general, boundary value problems allow for multiple solutions or no solutions at all. Consider for example the linear oscillator $\ddot x+x=0$ subject to the BC, $x(t_1=0)=0$ and $x(t_2=2\pi)=0$. There is an infinite number of solutions: $x(t)=A\sin(t)$ for any $A$. There are also BC where the 
action has no extremum orbit. An example is  $x(0)=0$ and $x(2\pi)=1$. There is no physical orbit which satisfies these BC. In this case, it is easy to see that action can acquire infinitely large (positive and negative) values for 
certain choices of the orbits. 
For  mixed BC where one of the conditions is $p=0$, the situation can even be more intriguing. 
The oscillator equation of motion constrained to  $\dot x(0)=0 $ and $x(2\pi)=1$, is solved by 
$x(t)=\cos(t)$. Let us compute the action $S=\int_0^{2\pi}dt ({\dot x}^2-x^2)/2 $ for the following choice for the perturbed orbits:  
$x_p=\cos(t)+A \cos(w t)$. These orbits satisfy the BC for $w=(2 n+1)/4$, but not the equation of motion. 
It is easy to see that $S=A^2\pi(w^2-1)/2$, i.e.  the extremum point is a maximum for $w<1$ 
and a minimum for $w>1$.

\subsection{Application to the Local Group (LG) of galaxies: masses of MW \& M31}
\begin{figure}[b]
 \vspace*{0cm}
\begin{center}
 \includegraphics[width=5.in]{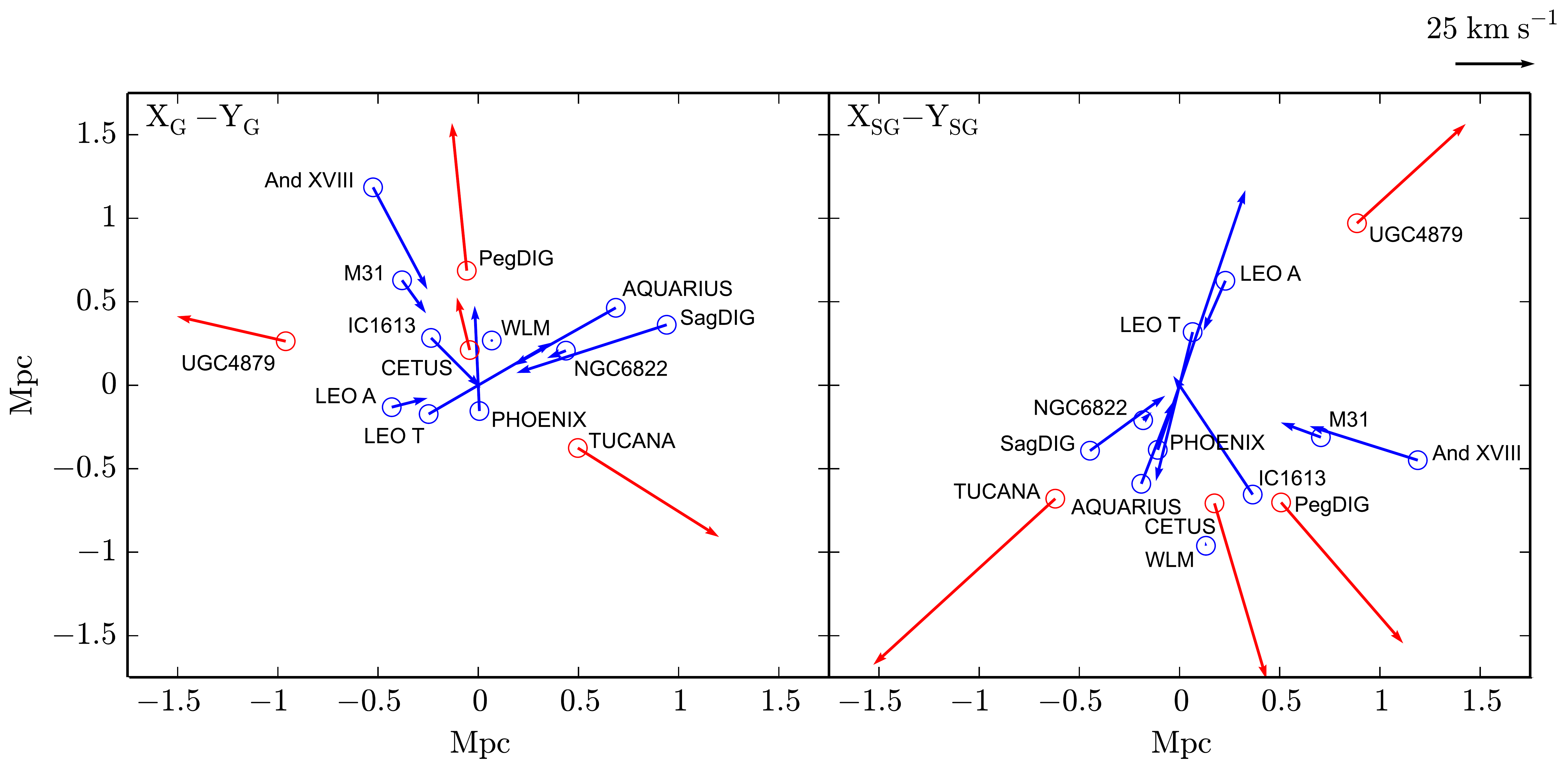} 
 \vspace*{0 cm}
 \caption{Galaxies in the LG in the Galactic (left) and SG (right) planes 
 the arrow represent observed radial velocities transformed to the frame of reference comoving with the 
 LG.  Curtsey of Ziv Mikulsy. }
   \label{fig3}
\end{center}
\end{figure}
The LG contains about  a dozen known  galaxies within a distance  $\sim 1.5$ Mpc. Although not a virialized 
object (see below), it is gravitationally bound and detached from the expansion. 
Galaxy members (excluding satellites) of the LG are shown in Fig. \ref{fig3}. The MW and M31 are by far the most luminous  (M31 is  100 more luminous than NGC6822-the third most luminous galaxy shown in the figure).
The  radial velocities in the LG frame are represented by the arrow in the figure. The (radial) velocity dispersion 
is $\sim 60 {\rm km s^{-1}}$ and  the flow pattern clearly reveals a non-virialized system that is  most likely is on a first 
 infall. Thus, the  virial theorem will  over-estimate the total mass of the LG. 
To  constraint the masses   of   MW and M31, we 
 apply the action principle to the nearby galaxies.
The  application is basically a generalization of the timing argue (TA) of Woltjer and Kahn who considered 
only the  MW and M31. By treating the two galaxies as point particles with zero angular momentum (relative to the center of mass),  
TA constrains the total mass $M_{MW}+M_{31}$ by demanding that the two galaxies originated from zero separation 
a Hubble time ago, reaching their current separation and relative velocity today. 
The action method breaks the degeneracy between the masses by  including  kinematical observations of the smaller members of the LG. Although dynamically unimportant, the observed  distances and velocities 
of these galaxies will allow us to resolve the individual masses  $M_{MW}$ and $M_{31}$. 

\begin{figure}[b]
 \vspace*{0cm}
\begin{center}
 \includegraphics[width=5.in]{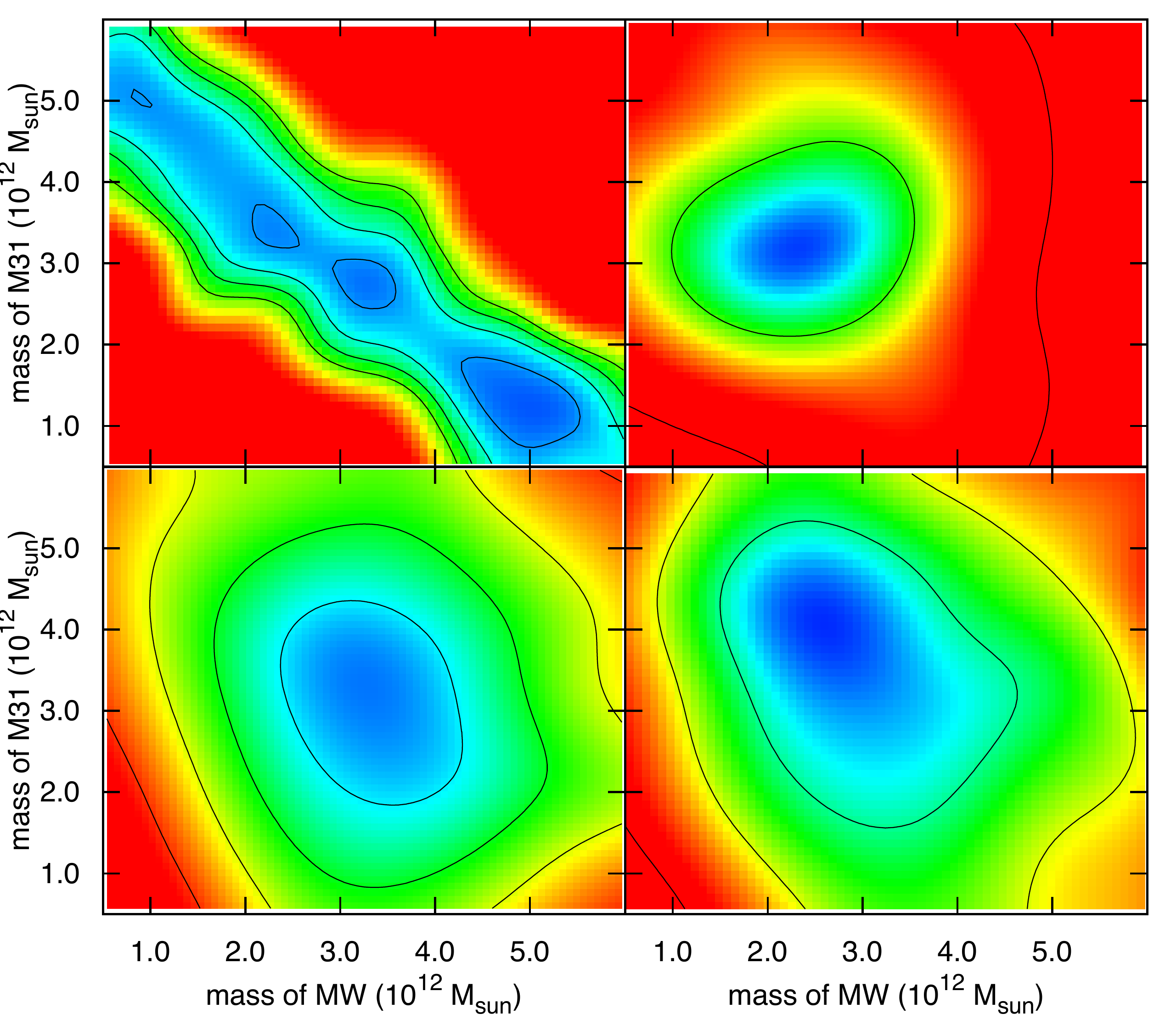} 
 \vspace*{0 cm}
 \caption{  Contours in $\chi^2$ for different values of MW  and M31  masses. Upper left: results from the two-body problem of MW + M31.  Upper right: LG actors only.  Lower left: LG actors + four external groups.  Lower right: same as lower left, with transverse velocity constraints included for five nearby galaxies.  The first contour level (solid black line) marks the region of 95\% confidence. From \cite{phelps}.}
   \label{fig4}
\end{center}
\end{figure}

 Galaxy orbits  which render the action stationary are found iteratively using standard techniques.
 They  are verified 
as solutions to   the EoM  in a leapfrog approximation.  
Since the solutions are non-unique, different choices of initial trial orbits will generally  give different solutions for the galaxy paths. We define a  $\chi^2$ measure of fit for all relevant observables, 
from which a best-fit solution can be selected.
Maps of $\chi^2$  in four different scenarios, all with $H_0 = 67$ and $\Omega_0 = 0.27$, are shown in Fig. \ref{fig4}.   At upper left are the contours in $\chi^2$ generated from a simplified catalog consisting of only MW and M31, to check the consistency of the action method  with the Timing Argument.  As expected, we find a well-defined constraint on the sum $M_{MW} + M_{M31}$. 
The  upper right panel, shows the results from a reduced version of our catalog which includes the LG actors but excludes external galaxies.  The additional  dynamical actors has broken the degeneracy in the TA, giving independent masses of $2.5 \pm 1.5 \times 10^{12} M_\odot$ for the MW and $3.5 \pm 1.0 \times 10^{12} \msun$ for M31.  With the addition of the external galaxies (Fig. \ref{fig4}, lower left), the best mass for the MW increases to $3.5 \pm 1.0 \times 10^{12} \msun$.  This is consistent at the lower end with previous TA measurements of the total LG mass and the individual MW mass.  When the transverse velocity constraints on M31, LMC, M33, IC10, and LeoI are added (lower right), the confidence intervals are broadened and the best-fit mass for MW decreases slightly, to $3.0 \pm 1.5 \times 10^{12} \msun$, reflecting the fact that lower masses for MW are correlated to lower transverse velocities for M31 and other nearby galaxies. 
These values are to be understood as the masses contained within roughly half  the separation between the 
two galaxies. For the MW, the value is more than twice what stellar motions yield for its virial mass. 
This could pose a challenge to the standard model since such an increase of the mass is not seen in N-body simulations.

\section{Cosmological constraints on 20-100 Mpc scales}

Restricting the analysis to these scales, away from non-linearities and hydrodynamical effects, greatly simplifies observational analyses of observations. Linear theory dynamical relations are, by and large, adequate on these scales but with 
the inclusion of scatter as a result from the presence of small scale modes.   
  Further,  large scale galaxy biasing is well described by a linear relation, as seen in Fig. \ref{fig2}. 

\subsection{Velocity-velocity comparison}

\cite{DN10}  have shown that GI passes a very important test: an excellent agreement 
between the velocity field predicted from the distribution of galaxies via linear theory and 
the observed motions of galaxies  obtained from the TF measurements of spiral galaxies. The beauty of this test is that the  effects of cosmic variance are minimized (in principles two good measurements are enough). 
The  contribution of Marc Davis to these proceedings offers more details.  

\subsection{The recovery of the CMB dipole, i.e. the motion of the LG with respect to the CMB} 

 The CMB temperature dipole  together with   astronomical estimation of the LG motion relative to the Sun,
provide  $V_{\rm lg}=627\pm 22\kms$ toward $(l,b)=(276^\circ\pm 3^\circ,30^\circ\pm 3^\circ)$  for 
 the LG motion relative to the CMB frame. The LG is accelerated by   
the cumulative gravitational pull of the surrounding large scale structure. Therefore,  an important probe of the GI paradigm is  whether the observed large scale structure, as traced by the galaxy distribution, 
could indeed account for the LG motion.  To do that, we need to compute the gravitational force on the LG
from an all sky survey of galaxies. According to  linear theory, this force should be proportional to the LG motion. 
We use the 
2MRS  which is 
 the deepest nearly-all sky survey of angular positions and spectroscopic galaxy redshifts limited to $K_s=11.75$
 and arguably the best sample of objects to estimate the LG motion.

A source of  uncertainty is  the  error in the linear theory dynamical reconstruction of the velocity from a given density field (equation \ref{eq:lin}).
To assess this error   we have estimated 
the LG motion from the full dark matter out to 
the largest possible outer  radius in the simulation, i.e. $R_{\rm out}=250\hmpc$. 
The corresponding $1\sigma$ error, $\sim 90\kms$, is substantially  smaller that the typical error  in linear reconstruction 
of the peculiar velocity of a generic observer in the Universe.
The reason for this is the strict criteria we 
have applied in  selecting the "LG observer" in the mock catalogs, aimed at  matching   the quietness and moderate density  environment of the 
observed LG.   Removing these selection criteria boosts the  error to $\gtsim 300 \kms$, consistent with previous studies \citep{NB00}.
Nonlinear dynamical reconstruction methods 
 can potentially reduce the dynamical error.
 However, because the particular environment of the LG,  errors due to linear reconstruction
 are subdominant compared to the total error budget.  
 
\begin{figure}[b]
 \vspace*{0cm}
\begin{center}
 \includegraphics[width=3.5in]{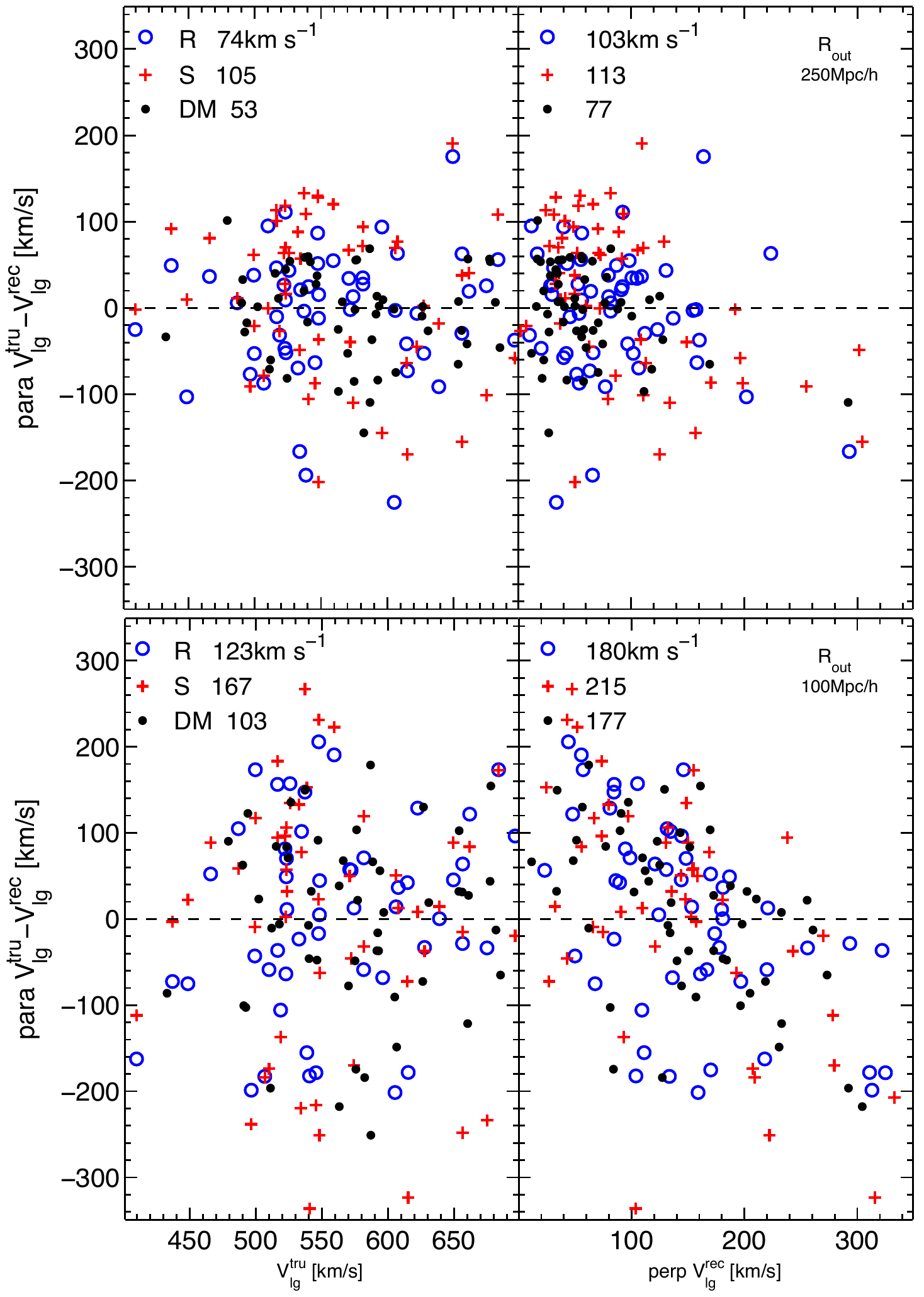} 
 \vspace*{0 cm}
 \caption{A scatter plot showing  the velocity residual in the parallel and perpendicular directions  from mock catalogs. 
The blue dots are in real space and the red crosses are in  redshift space, while black dots show recovery from the full dark matter density field in real space. 
The  {\it rms} values of the  parallel and perpendicular residuals are listed in the left and right panels, respectively.
  Top and bottom panels correspond to velocity reconstruction with only data  within $R_{\rm out}=250\hmpc$
 and $R_{\rm out}=100\hmpc$, respectively.
The  {\it rms}  of the parallel and perpendicular residuals are indicated, respectively,  in the left and right panels. 
  }
   \label{fig5}
\end{center}
\end{figure}

 For a more realistic  assessment of the recovery of the LG motion in real data we resort to  mock catalogs designed to match the  
2MRS with $K_s< 11.75$ (as in Fig. \ref{fig3}). 
The results for various cases are shown in Fig. \ref{fig5}, as described in the caption. There is a dramatic 
decrease in the scatter when matter within $R_{\rm out}=250\hmpc$ is included. Still the residuals are at the 
level of $70-100\kms$ depending on the case considered. 
This is consistent with \cite{bilicki11} who derived a similar result analytically. 
Current all sky data do not allow a reliable assessment of the contribution of fluctuations beyond $100\hmpc$. Much 
of that is because of the Kaiser rocket effect \citep{k87} (see \cite{NDBLG} for further details).

\subsection{The bulk flow} 
The bulk flow, $\vB(r)$, is the mean motion of a sphere of a given radius, $r$. 
Computing  $\vB$ from the  sparse and noisy peculiar velocity catalogs is very challenging.
Improper handling of the data may easily lead to artificially large flow. 
\cite{ND11} have estimated $\vB(r) $ from   the  SFI++ survey.   They have discarded   $a)$ the fainter galaxies which do not obey the TF relation and $b)$  galaxies beyond 100$\hmpc$ which are likely to suffer from systematics related to the measurements of TF parameters. 
They found $\vB(r)$  to be consistent 
with the standard $\Lambda$CDM model and derived the constraint:  
$\sigma_8(\Omega_m/0.266)^0.28=0.86\pm 0.11$,  which leads to a $\sigma_8$ higher (although consistent with) 
than the WMAP value. However, taking  $\Omega_m=0.317$ from the recent Planck results, yields a best fit value $\sigma_8=0.819$, very close to the result reported by the Planck collaboration.
Despite earlier claims of anomalous bulk flows,  the emergent consensus  is that 
 bulk flow measurements  from numerous data (intrinsic relations, SN and kSZ),  is consistent with the standard $\Lambda$CDM model 
 \citep{ND11,Dai11,2011MNRAS.414..264C,2014MNRAS.445..402H,Feix14,2014arXiv1407.6940W,2014MNRAS.437.1996M,2014A&A...561A..97P}.

%

\section{Summary}

The gravitational instability model for structure formation with its current $\Lambda$CDM incarnation
described the nearby large scale structure very well. 
Whatever corrections for this model should be small as far as large scales are concerned. 
Probing deviations from this model on large scales maybe possible only with next generation redshift surveys. 
Modifications on smaller scales ($\sim $ a few Mpcs) are a different matter \citep[e.g.][]{PN10} but are not the subject of this contribution. 

Upcoming redshift surveys will contain a large number of galaxies to allow 
constraints on  the cosmological velocity field independently of the classical redshift distortion 
 the correlation functions. 
The idea is that galaxy redshifts $cz$ depend on $v_p$. Hence, using $cz$  instead of true distances, $d$,  in order to estimate galaxy luminosities (from the observed apparent magnitudes) will introduce spatial variations of the luminosity function. Assuming negligible environmental dependence in the luminosity function, these variations 
can put constraints on the velocity field. 
\cite[e.g.][]{1980ApJ...242..448Y, NBETA, Feix14}.  Basically this method 
assumes that galaxy luminosity is a standard candle where the very large distance error is beaten by the large number of galaxies. The method can be applied to surveys with photometric  redshifts \citep{2MPZ} and is not restricted to
spectroscopic surveys. 
The method can  constrain galaxy biasing and the the linear growth rate is a velocity model based on 
the actual galaxy distribution is used to model the luminosity variations.
The contribution of Martin Feix addresses potential caveats and presents an application to the SDSS.

Another potential probe is offered by Gaia. There could be a large number of galaxies detected  as point sources 
by Gaia \citep{NGAIA}. For example, the nuclei of M87 and N5121 (both at d=17.8Mpc) should be detected with an end of mission accuracy of $600\kms$ in the transverse motion. The surface brightness profiles of the Carnegie-Irvine Galaxy Survery show that 70\% of galaxies in this survey could be detected by Gaia. The majority of these nearby galaxies will be detected if placed at $\gtsim 500 \rm Mpc$ (early type) and $\gtsim \rm 250 Mpc$ (late type).
Of course, parallax distance errors increase quadratically with distance. Therefore, the error on Gaia's  distances for 
extragalactic objects will be huge  and cannot be used to get the transverse velocities from the measured proper motions. But, at such distances, the redshifts can be used as  proxies for the true distance without introducing a significant error  in the transverse velocities.



\end{document}